\documentclass[twocolumn,showpacs,preprintnumbers,amsmath,amssymb,prl,footinbib]{revtex4-1}
\usepackage{graphicx}
\usepackage{dcolumn}
\usepackage{bm}
\usepackage[tight]{subfigure}
\usepackage{psfrag}
\usepackage{verbatim}
\usepackage[usenames,dvipsnames]{color}
\usepackage{xspace}

\newcommand{\Q}{pumped current\xspace}
\newcommand{\Fig}[1]{Fig.~\ref{#1}}
\newcommand{\Eq}[1]{Eq.~(\ref{#1})}
\newcommand{\eq}[1]{(\ref{#1})}
\newcommand{\Ref}[1]{Ref.~\cite{#1}}
\newcommand{\half}{\textstyle{\frac{1}{2}}}
\renewcommand{\vec}[1]{\mathbf{#1}}
\newcommand{\ket}[1]{|{#1}\rangle}

\newcommand{\didv}{\ensuremath{d\bar{I}_\mathrm{L}^{(\mathrm{i})} / d\bar{V}_\mathrm{b}}\xspace}
\newcommand{\Ia}{\ensuremath{\bar{I}^{(\mathrm{a})}_{r}}\xspace}
\newcommand{\IaL}{\ensuremath{\bar{I}^{(\mathrm{a})}_{\mathrm{L}}}\xspace}
\newcommand{\Ii}{\ensuremath{\bar{I}^{(\mathrm{i})}_{r}}\xspace}
\newcommand{\Iat}{\ensuremath{I^{(\mathrm{a})}_{t,r}}\xspace}
\newcommand{\Iit}{\ensuremath{I^{(\mathrm{i})}_{t,r}}\xspace}

\begin{document}
\title{
Interaction-induced adiabatic non-linear transport
}
\author{Felix Reckermann$^{(1,3)}$}
\author{Janine Splettstoesser$^{(2,3)}$}
\author{Maarten R. Wegewijs$^{(1,2,3)}$}
\affiliation{
  (1) Institut f\"ur Festk{\"o}rper-Forschung - Theorie 3,
      Forschungszentrum J{\"u}lich, 52425 J{\"u}lich,  Germany \\
  (2) Institut f\"ur Theoretische Physik A,
      RWTH Aachen University, 52056 Aachen,  Germany \\
  (3) JARA- Fundamentals of Future Information Technology\\
}
\begin{abstract}
We calculate the time-dependent non-linear transport current through an interacting quantum dot in the single-electron tunneling regime (SET). We show that an additional dc current is generated by the electron-electron interaction by adiabatic out-of-phase modulation of the gate and bias voltage. This current can arise only when two SET resonance conditions are simultaneously satisfied. We propose an adiabatic transport spectroscopy where lock-in measurement of a ``time-averaged stability diagram'' probes interactions, tunnel asymmetries and changes in the ground state spin-degeneracy.
\end{abstract}
\maketitle
{\em Introduction.}
Transport through nano-scale devices modulated by time-dependent externally applied electric fields
is an active field of research important for
transport spectroscopy and manipulation of the charge and spin degrees of freedom in nano-structures, see, e.g.,~\cite{Hanson07rev}.
A particularly gentle way of time-dependently probing a system
is through ``adiabatic pumping''~\cite{Brouwer98,*Buettiker94,*Moskalets02b,Zhou99,*Makhlin01,*Pothier92,*Switkes99,Entin02}.
Here a finite dc current is generated in the absence of an applied bias
by a weak, low frequency periodic modulation of system parameters.
Adiabaticity 
in a transport situation means that
many electrons visit the system during one cycle of the driving with frequency $\Omega$
and that the modulation is too weak and too slow to excite the device by direct state-to-state transitions.
The current generated this way crucially depends on
which subset of parameters is modulated,
on the working point about which the modulation takes place
and on interactions, which are of special importance in nano-scale devices.
For strictly adiabatic pumping one needs to vary at least two parameters,
single parameter pumping requiring a higher frequency~\cite{Grifoni98}.
Among the various combinations of parameters studied so far,
the modulation of the applied bias has received little attention~\cite{Moskalets04}.
Most works have considered small deviations around an \emph{equilibrium} working point
where no steady state current is flowing.
Adiabatic modulation around a non-equilibrium transport state induced by a static non-linear bias voltage
has been explored only
for systems with negligible Coulomb interaction~\cite{Entin02},
motivated by experiments with surface acoustic waves~\cite{Fletcher03,*Robinson02}.
Non-linear bias voltage and Coulomb interaction have received little theoretical attention  in the adiabatic regime.
Limited to an equilibrium working point, some works have studied interacting quantum dots~\cite{Aleiner98,*Aono04,*Cota05a,*Splettstoesser05,*Sela06,*Fioretto08,*Arrachea08,Splettstoesser06}
and wires~\cite{Citro03,*Das05}.
Including the effect of strong interactions beyond the mean-field picture is a challenge,
since the powerful scattering matrix approach~\cite{Brouwer98,*Buettiker94,*Moskalets02b} breaks down here.
Generally, one expects the additional non-equilibrium introduced by a static dc-bias voltage,
in combination with strong electron-electron interactions, to strongly modify the pumping,
 providing novel opportunities to investigate and control transport properties of nano-scale devices.
 
In this letter we propose a new scheme for transport spectroscopy of interacting systems using adiabatically time-dependent electric fields.
We analyze an interacting quantum dot in the SET regime,
adiabatically driven by out-of-phase gate and bias potentials.
In contrast to previous works, the applied bias can be arbitrary,
i.e., we modulate the parameters around a steady non-equilibrium state supporting a finite dc current.
We show that the strong local interaction \emph{generates} an additional adiabatic dc current,
which is identically zero without interaction for any value of the applied voltages and magnetic field.
We propose to use this effect as a tool for non-linear transport spectroscopy
which can be measured using lock-in techniques.
The adiabatic dc current
is non-zero only when two conditions for single-electron tunneling are simultaneously satisfied.
Plotted as function of the time-averaged gate and bias voltage,
it gives rise to a new type of ``stability diagram''.
Furthermore, we show that in an external magnetic field lifting the spin-degeneracy,
the adiabatic modulation only gives rise to transport effects in the regime of non-linear bias,
which qualitatively distinguish between different junction asymmetries.
 
\begin{figure}[htbp]
  \begin{tabular}{cc}
    \includegraphics[width=0.45\linewidth]{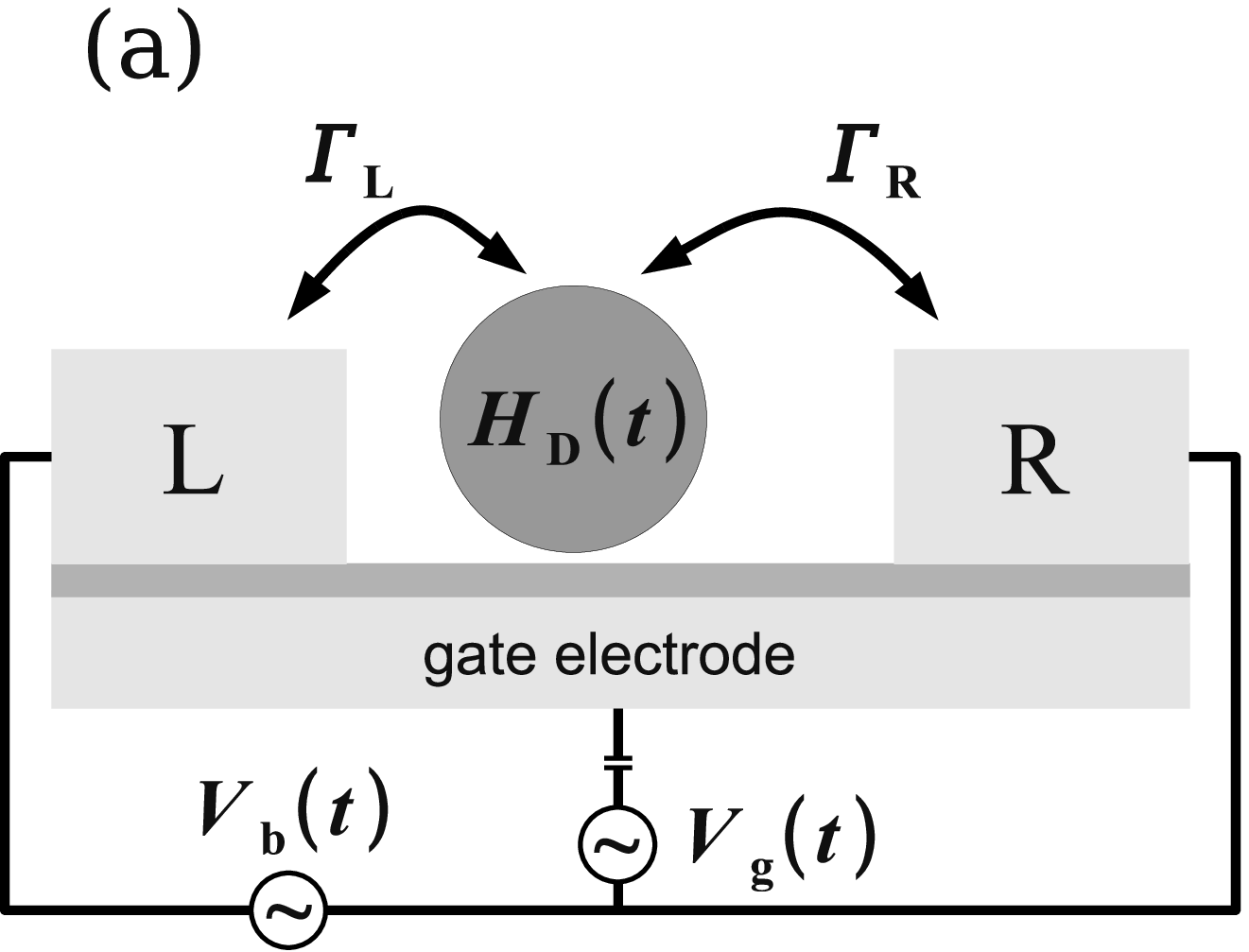}  &
    \includegraphics[width=0.48\linewidth]{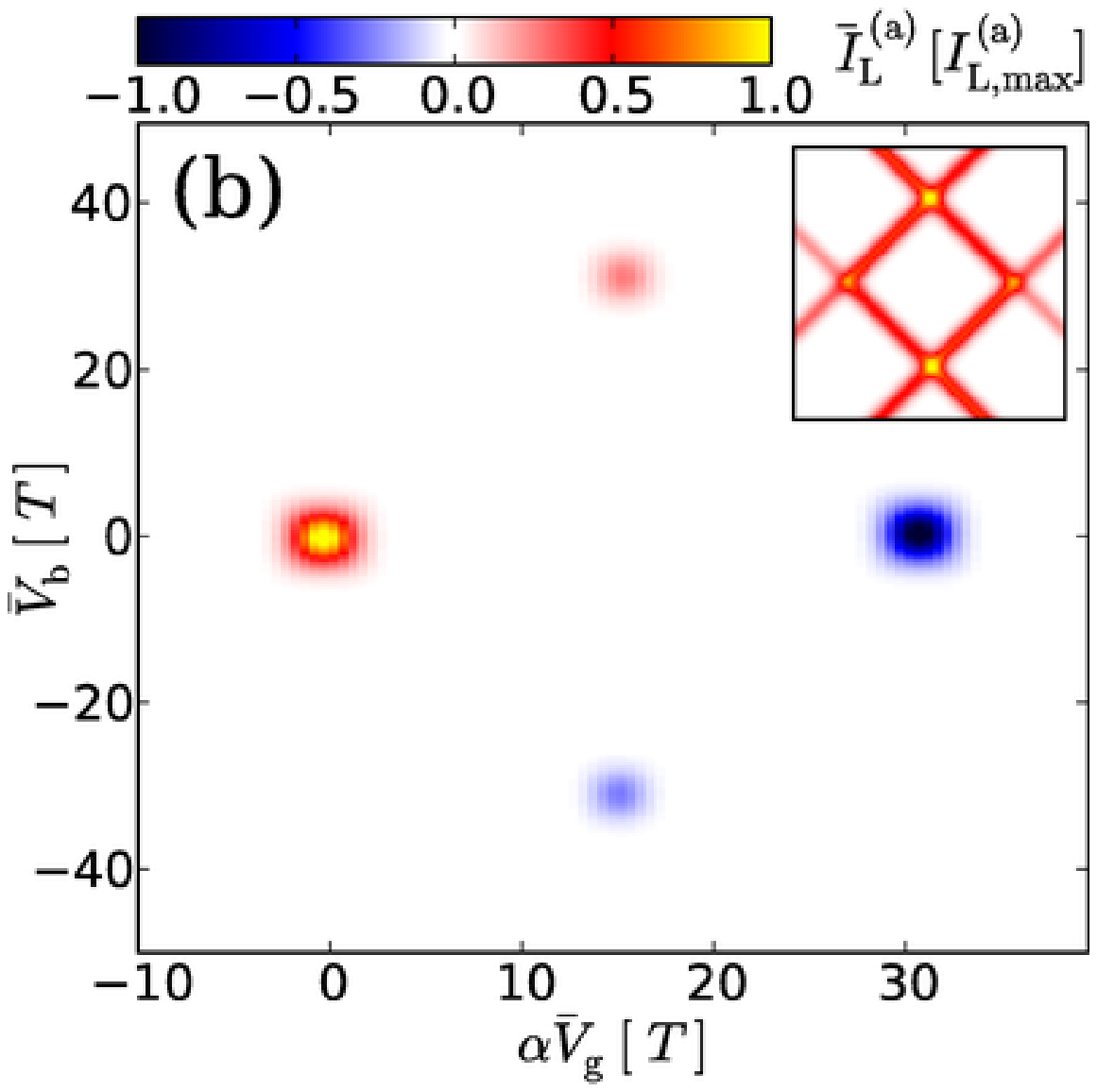}\\
    \includegraphics[width=0.48\linewidth]{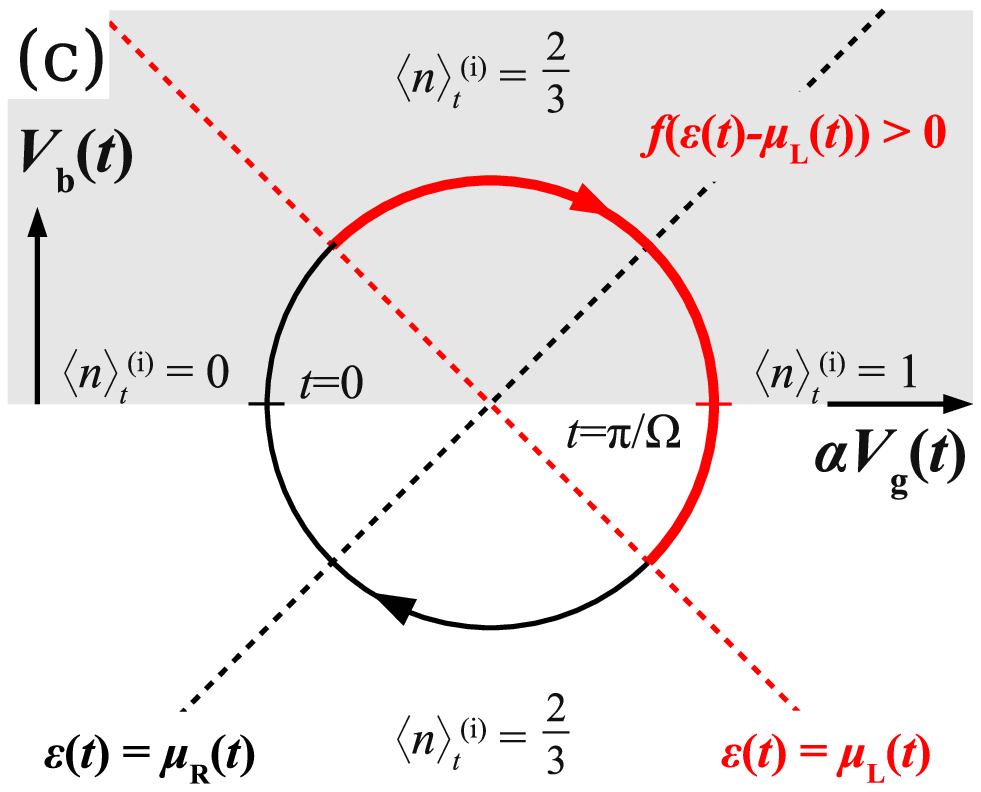} &
    \includegraphics[width=0.48\linewidth]{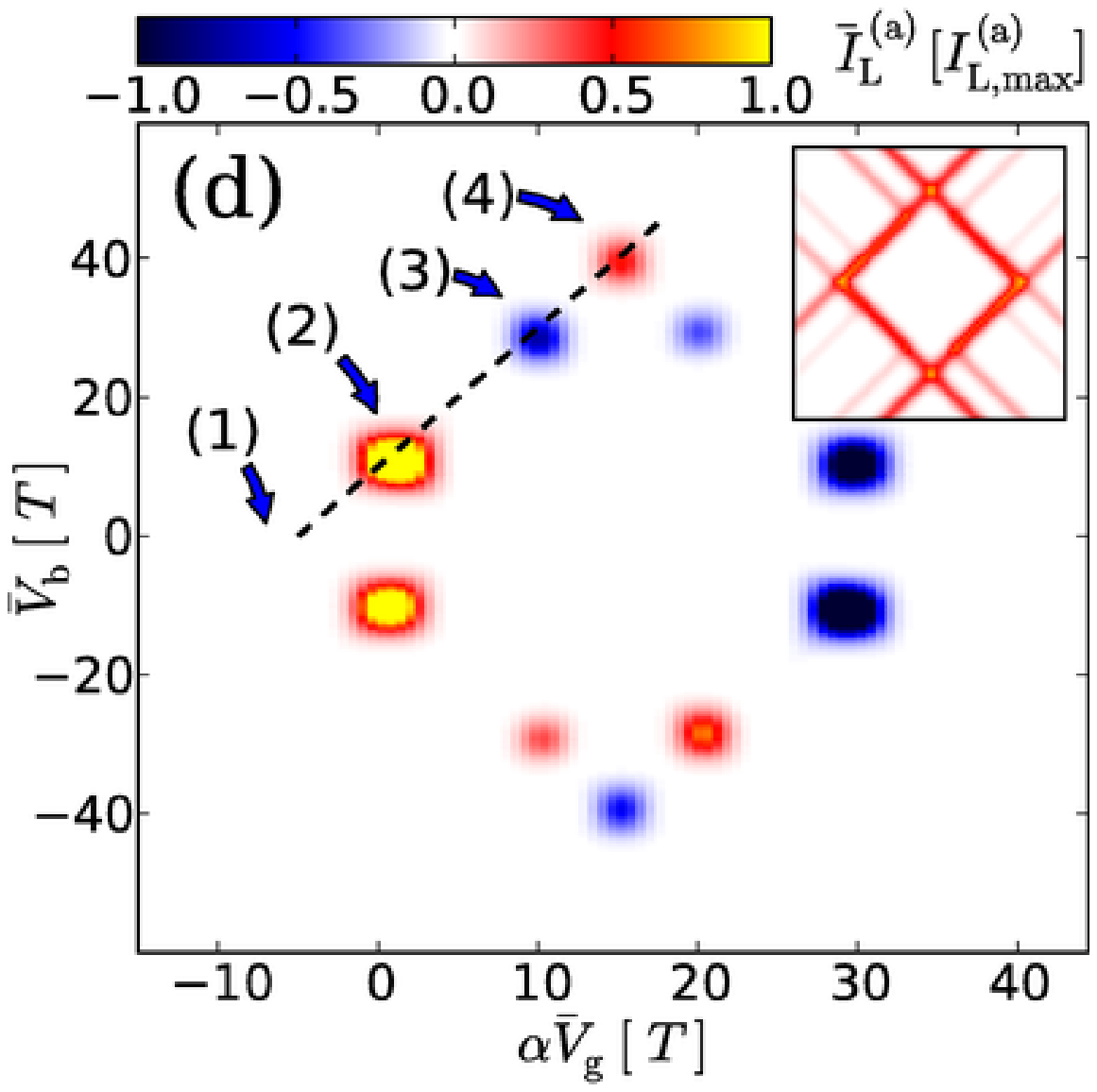}
  \end{tabular}
  \caption{
    (Color online).
    (a) Sketch of the transport setup.
    (b) Adiabatic dc current through the left junction, \IaL,
    as function of the time-averaged gate and bias voltage for Coulomb interaction $U=30 T$,
    zero magnetic field and $\Gamma=0.5 T$, $\lambda=0.25$ and driving parameters
    $\Omega=0.1 T$, $\alpha \delta V_\mathrm{g} = \delta V_\mathrm{b} = 0.5 T$.
    We plot \IaL scaled to the maximal absolute value at the degeneracy points,
    $\bar{I}^{\mathrm{(a)}}_\mathrm{max,L} = \frac{2\Omega}{27}\frac{\Gamma_\mathrm{L}\Gamma_\mathrm{R}}{(\Gamma/2)^2}  \frac{\alpha \delta V_\mathrm{g}}{4T} \frac{\delta V_\mathrm{b}}{4T}$,
    consisting of the frequency, a coupling asymmetry factor, and the ratio of the voltage phase-space factors to the thermal energy window.
    Inset: differential conductance, \didv, versus
    $\bar{V}_\mathrm{g}$ and $\bar{V}_\mathrm{b}$.
    (c) Voltage modulation cycle around the degeneracy point $(\alpha \bar{V}_\mathrm{g},\bar{V}_\mathrm{b})\approx (0,0)$
    with dashed resonance lines $\varepsilon(t) = \mu_r(t)$.
    (d) Same as (b) for finite applied magnetic field, $B=10 T$.
  }
  \label{fig:1}
\end{figure}

{\em Model.}
We consider a quantum dot weakly coupled to two electrodes as sketched in \Fig{fig:1}(a).
The gate and bias voltage are modulated with frequency $\Omega$ around the working point specified by the voltages $ \bar{V}_\mathrm{g}$ and $\bar{V}_\mathrm{b}$:
\begin{align}
  V_x(t)=\bar{V}_x+\delta V_x \sin(\Omega t + \varphi_x), \qquad x=\mathrm{b},\mathrm{g}.
\end{align}
We consider the important case where a single orbital level with strong Coulomb interaction $U$
is relevant for transport.
We denote the spin-resolved dot number operator by $n_\sigma=d_{\sigma}^{\dagger}d_{\sigma}$,
where the spin $\sigma=\uparrow,\downarrow$ is quantized along the external magnetic field (if present).
The Hamiltonian reads
$H_{\text{D}}(t)=\sum_\sigma \epsilon_\sigma(t) n_\sigma + U n_\uparrow n_\downarrow$.
The energy of an electron created by $d_{\sigma}^{\dagger}$ equals
$\epsilon_\sigma(t)= -\alpha V_\mathrm{g}(t) + \sigma B/2$
using the shorthand $\sigma=\pm1$ for spin $\uparrow,\downarrow$.
Importantly, the time-dependent gate voltage $V_\mathrm{g}(t)$ capacitively modulates this energy with lever arm $\alpha < 1$.
Furthermore,  $B$ is the Zeeman energy in units $e=\hbar=k_\mathrm{B}=1$.
The many-body eigenstates of $H_D(t)$ are
$\ket{0}$,
$\ket{\sigma} = d^{\dagger}_{\sigma}                        \ket{0}$ with $\sigma=\uparrow,\downarrow$ and
$\ket{2}      = d^{\dagger}_{\uparrow} d^{\dagger}_{\downarrow} \ket{0}$
with energies $0$, $\epsilon_\sigma(t)$, $\sum_\sigma \epsilon_\sigma(t)+U$, respectively.
The time-dependent bias  $V_\mathrm{b}(t)$ enters through the electro-chemical potentials $\mu_r(t)=\pm V_\mathrm{b}(t)/2$ of electrodes $r=\mathrm{L},\mathrm{R}$, which are described by
$H_{\text{E}}(t)=\sum_{r,k,\sigma} \left( \epsilon_{k}+\mu_r(t)\right) c_{rk\sigma}^{\dagger} c_{rk\sigma}$.
Finally, $H_{\text{T}}=\sum_{r,k,\sigma} t_r c_{rk\sigma} d_{\sigma}^{\dagger}+H.c.$ describes the tunneling between the dot and the electrodes, with tunnel coupling strength $\Gamma_r=2\pi \rho_r |t_r|^2$, where $t_r$ is the amplitude and $\rho_r$ the density of states of lead $r=\mathrm{L,R}$.
We define $\Gamma=\sum_r \Gamma_r$.
We consider here the important case where the transport is affected dominantly through the modulation of the energy level positions $\varepsilon_\sigma(t)$ and the bias energy window $\mu_{\mathrm{L}}(t) - \mu_{\mathrm{R}}(t)$.
The change in the tunnel coupling is of negligible importance.
This is the typical situation in non-linear transport spectroscopy of quantum dots.
It holds in particular for small amplitude modulation of the voltages considered here.

{\em Retarded occupations and transport current.}
The total Hamiltonian
$ H(t) = H_{\text{E}}(t) + H_{\text{D}}(t) + H_{\text{T}}$
thus contains strong  interaction on the dot,
an adiabatic time-dependence
and non-equilibrium introduced by the non-linear bias voltage.
Within the framework of the real-time transport theory
the time-dependent occupation probabilities of the many-body dot states,
$\vec{p}(t)=(p_0(t),p_\downarrow(t) , p_\uparrow(t), p_2(t))$
can be shown to obey the kinetic equation~\cite{Splettstoesser06}
\begin{align}
  \dot{\vec{p}}(t) = &~\int_{-\infty}^t dt' \vec{W}(t,t') \vec{p}(t') \ .
  \label{eq:general_master}
\end{align}
The kernel, $\vec{W}(t,t')$, accounts for changes of the dot occupations due to electron tunnel processes to/from the electrodes.
Although,  it explicitly depends on both time arguments $t$ and $t'$ (in contrast to the time-independent case)
it can be calculated perturbatively for slowly varying fields~\cite{Splettstoesser06}.
Here we restrict ourselves to the lowest order contributions in both the tunneling coupling
 (single-electron tunneling (SET))
and in the time-dependent perturbation of external system parameters (adiabatic driving).
We consistently solve the kinetic equation by expanding it around the instantaneous reference solution, $\vec{p}_t^{(\mathrm{i})}$, defined by
\begin{align}
    0 = & ~\vec{W}_t^{(\mathrm{i})} \vec{p}_t^{(\mathrm{i})}.
    \label{eq:instantaneous_kinetic_equation}
\end{align} 
The instantaneous kernel and its zero-frequency Laplace transform
$\vec{W}^{(\mathrm{i})}_t= \lim_{\eta \downarrow 0}  \int_{-\infty}^t dt' \vec{W}^{(\mathrm{i})}(t-t')e^{\eta(t-t')}$
are evaluated using stationary transport theory in lowest order in $\Gamma$.
In this limit of weak coupling, $\Gamma \ll T$, where $T$ is the electron temperature,
the result reduces to Fermi's Golden Rule.
In \Eq{eq:instantaneous_kinetic_equation} the voltages are replaced by their instantaneous values at measuring time, $V_x \rightarrow V_x(t)$, $x=\mathrm{b},\mathrm{g}$,
resulting in a parametric time-dependence indicated by the subscript $t$.
Hence, $\vec{p}_t^{(\mathrm{i})}$ would be the \emph{time-dependent steady state} of the dot if the system was able to follow the parameter modulation instantaneously.
By inserting $\vec{p}(t) = \vec{p}_t^{(\mathrm{i})} +\vec{p}_t^{(\mathrm{a})}$ in \Eq{eq:general_master}, using~\Eq{eq:instantaneous_kinetic_equation},
and doing a systematic lowest order expansion in $\Omega / \Gamma \ll 1$,
we find for the first correction $\vec{p}_t^{(\mathrm{a})}$ to the instantaneous reference solution $\vec{p}_t^{(\mathrm{i})}$:
\begin{align}
  \dot{\vec{p}}_t^{(\mathrm{i})} = &~ \vec{W}_t^{(\mathrm{i})} \vec{p}_t^{(\mathrm{a})}
  \label{eq:adiabatic_kinetic_equation}.
\end{align}
This correction accounts for the actual delay suffered by the system
due to the finite rate of sweeping the voltages.
Further corrections to this adiabatic approximation can be neglected if in addition
$\alpha \delta V_\mathrm{g},
 \delta V_\mathrm{b} \ll T^2/\Omega$.
The time-dependent steady state including the retardation is uniquely determined
by equations \eq{eq:instantaneous_kinetic_equation} and \eq{eq:adiabatic_kinetic_equation},
together with the normalization conditions
$\vec{e}^{\text{T}}\vec{p}_t^{(\mathrm{i})}=1$ and
$\vec{e}^{\text{T}}\vec{p}_t^{(\mathrm{a})}=0$ with $\vec{e}^{\text{T}}=\left(1,1,1,1\right)$.
 
The time-dependent current flowing from lead $r=\mathrm{L},\mathrm{R}$ into the dot is found in a similar way~\cite{Splettstoesser06}
and can be decomposed into two corresponding parts,
\Iit and \Iat.
Here \Iat is the adiabatic correction to the current due to the retardation of the system, 
i.e., it vanishes in the limit $\delta V \Omega \rightarrow 0$.
The central quantities discussed in this paper are obtained when averaging the two current contributions over one entire modulation cycle
$\bar{I}_r^{(\mathrm{i/a})}  =  \frac{\Omega}{2\pi} \int_0^{\frac{2\pi}{\Omega}} dt ~ I_{t,r}^{(\mathrm{i/a})}$.
Here, \Ii equals the dc current one would measure for time-independent voltages equal to $\bar{V}_\mathrm{g}$ and $\bar{V}_\mathrm{b}$.
Plotting \didv as function of these voltages, one obtains the standard Coulomb blockade stability diagram~\cite{Hanson07rev}.
The quantity of central interest here, \Ia, is the additional dc current component
due to the retardation of the quantum dot state.
This quantity can be obtained, e.g., by subtracting from the total measured time-averaged current its zero frequency limit.
 
For the time-dependent adiabatic current, we obtain a central analytic result
\begin{align}
  \Iat & =
  \frac{(\Gamma_r+\gamma_r) (\Gamma-\gamma) + \beta_r \beta}
  {\Gamma^2 -\gamma^2+\beta^2}
  ~\frac{d}{dt} \langle n \rangle_t^{(\mathrm{i})}
  \nonumber\\
  & + 2 
  \frac{(\Gamma_r+\gamma_r) \beta         - \beta_r (\Gamma+\gamma)}
  {\Gamma^2 -\gamma^2+\beta^2}  
  ~\frac{d}{dt} \langle S_z \rangle_t^{(\mathrm{i})}.
  \label{eq:Ia_finiteB}
\end{align}
The average instantaneous charge, 
$\langle n   \rangle_t^{(\mathrm{i})} = \sum_{\sigma}        p_{t,\sigma}^{(\mathrm{i})} + 2p_{t,2}^{(\mathrm{i})}$, 
and spin, 
$\langle S_z \rangle_t^{(\mathrm{i})} = \sum_{t,\sigma} (\sigma/2) p_{t,\sigma}^{(\mathrm{i})}$
are found from \Eq{eq:instantaneous_kinetic_equation}.
Although the time-dependent adiabatic currents depend on the junction $r$ where they are evaluated,
their time averages are related by charge conservation,
$\sum_r \Iat = \frac{d}{dt} \langle n \rangle_t^{(\mathrm{i})}$,
giving $\sum_r \Ia=0$.
The prefactors in \Eq{eq:Ia_finiteB} contain
\begin{align}
  \gamma_r(t) &= {\half} \Gamma_r \sum_\sigma        \left[f\left(\epsilon_{\sigma r}(t)  \right)
                                                 -f\left(\epsilon_{\sigma r}(t)+U \right) \right] \label{eq:gamma} \\
  \beta_r(t)  &= {\half} \Gamma_r \sum_\sigma \sigma  \left[f\left(\epsilon_{\sigma r}(t)  \right)
                                                 -f\left(\epsilon_{\sigma r}(t)+U \right) \right] \label{eq:beta}
\end{align}
and their sums by $\gamma=\sum_r \gamma_r$, and $\beta=\sum_r \beta_r$,
where $\sigma = \pm$ (corresponding to $\uparrow,\downarrow$) and $r=\mathrm{L,R}$.
All these quantities depend on time through the distance to resonance $\epsilon_{\sigma r}(t)=\epsilon_\sigma(t) - \mu_r(t)$
in the arguments of the Fermi-function $f(\omega)=(\exp(\omega/T)+1)^{-1}$.
From \Eq{eq:Ia_finiteB} we infer a necessary condition for a non-vanishing time-averaged adiabatic current which also holds for more complex systems:
\IaL can only be non-zero if two SET resonance conditions are satisfied \emph{simultaneously}.
If only a single resonance condition is satisfied (effectively this is single-parameter pumping), (\ref{eq:Ia_finiteB}) is a total time-derivative of a periodic function, resulting in a zero time-average.
The resonances in \IaL are thus located at resonance line crossings of the standard \didv map.

{\em Interaction-induced dc current.}
The central result of the paper relates to  the prefactors in \Eq{eq:Ia_finiteB}.
Since the tunnel rates $\Gamma_r$ and $\Gamma=\sum_r \Gamma_r$  are time-independent
it is clear that
the adiabatic dc current is generated by the Coulomb interaction $U$.
Indeed, since $\gamma_r=\beta_r=0$ for $U=0$ the adiabatic current
$\Iat=({\Gamma_r}/{\Gamma}) \left( {d\langle n \rangle_t^{(\mathrm{i})}}/{dt}+2{d\langle S_z \rangle_t^{(\mathrm{i})}} / {dt} \right)$
is a total time-derivative, which, integrated over a period, yields $\Ia=0$.
We emphasize that in this case the current \Ia vanishes identically for any value
of the time-averaged external voltages and
of the time-independent tunnel couplings and external magnetic field.
We now discuss the voltage dependence of the adiabatic dc current in the experimentally important regime of strong local interaction $U \gg T \gg \Gamma$.
We fix the direction of the modulation cycle by taking $\varphi_\mathrm{b}=0$ and $\varphi_\mathrm{g} = - \pi/2$,
for which the adiabatic dc current is maximal,
and time-average the current numerically.
We first focus on the case of zero magnetic field for which
$\varepsilon_\sigma(t)=\varepsilon(t)=-\alpha V_\mathrm{g}(t)$ is independent of spin $\sigma$.
Therefore $\beta_r=0$ and \Eq{eq:Ia_finiteB} simplifies to
$\Iat = ( \Gamma_r + \gamma_r ) /
                ( \Gamma   + \gamma   ) \, {d\langle n \rangle^{(\mathrm{i})}_t} / {dt}$.
In \Fig{fig:1}(b) we show a \emph{time-averaged} stability diagram, i.e., \IaL plotted as function of the time-averaged gate and bias voltage.
In contrast to the standard (\didv) stability diagram in the inset of \Fig{fig:1}(b)
this map of \Q indeed shows resonant enhancements only at discrete points of size $\propto T$ where two SET resonances meet.
Most prominent are the two charge degeneracy points $(\alpha \bar{V}_\mathrm{g},  \bar{V}_\mathrm{b}) \approx (0,0)$ and $(U,0)$ at which
the adiabatic dc current has opposite sign and maximum amplitude.
We now explain the microscopic origin of the positive sign of the adiabatic dc current at the degeneracy point $(\alpha \bar{V}_\mathrm{g}, \bar{V}_\mathrm{b}) \approx (0,0)$ for symmetric tunnel coupling $\Gamma_\mathrm{L}=\Gamma_\mathrm{R}$.
For $t \in (0,\pi/\Omega)$ the adiabatic current through the left junction is positive, $I_{t,\mathrm{L}}^{(a)} > 0$,
whereas in the second half of the cycle $I_{t,\mathrm{L}}^{(a)} < 0$.
This is because $\Gamma + \gamma   \approx \Gamma   ( 1 + \sum_r f(\varepsilon(t)-\mu_r(t))/2 )$ and
${d\langle n \rangle^{(\mathrm{i})}_t} / {dt}$ are symmetric and antisymmetric functions of the time $t$.
The time-average \IaL is nevertheless non-zero due to the factor
$\Gamma_\mathrm{L} + \gamma_\mathrm{L} \approx \Gamma/2 +  \Gamma f(\varepsilon(t)-\mu_\mathrm{L}(t))/2$ in the numerator.
Clearly, since the first term is constant, the non-zero time-average comes from the contribution $\propto f(\varepsilon(t)-\mu_\mathrm{L}(t))$
which is non-zero for times for which $\varepsilon(t) < \mu_\mathrm{L}(t)$ (red part of cycle in \Fig{fig:1}(c)).
One thus samples predominantly the loading parts of the cycle where ${d\langle n \rangle^{(\mathrm{i})}_t} / {dt} > 0$ (shaded in \Fig{fig:1}(c)),
where an excess of electrons tunnels onto the dot through the left junction.
Therefore the adiabatic dc current is positive.
Similarly, one finds for the point $(\alpha\bar{V}_\mathrm{g}$, $\bar{V}_\mathrm{b}) \approx (U,0)$ the opposite adiabatic dc current
due to the negative sign of the second term in \Eq{eq:gamma}.
For asymmetric rates $\Gamma_\mathrm{L} \neq \Gamma_\mathrm{R}$ the time-dependence of $\Gamma+\gamma$ becomes important as well,
but does not alter the sign of the adiabatic dc current.
\begin{figure}[t]
  \includegraphics[width=0.5\linewidth]{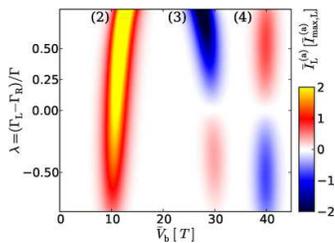}
  \caption{
    (Color online). 
    The \IaL along the line $\varepsilon = \mu_\mathrm{R}$
    (dashed line in \Fig{fig:1}(d)),
    as function of the tunnel coupling asymmetry parameter $\lambda$,
    all other parameters being the same as in \Fig{fig:1}(d)
    and $\bar{I}^{\mathrm{(a)}}_\mathrm{max,L}$ is taken for $\lambda=0$ ($\Gamma_\mathrm{L} = \Gamma_\mathrm{R}$).
  }
  \label{fig:2}
\end{figure}
In a magnetic field $B \gg T$  the adiabatic dc current plotted in \Fig{fig:1}(d) is completely suppressed
 in the linear response regime $\bar{V}_\mathrm{b} \ll T $.
Indeed, in this limit, $\gamma_r = \beta_r$ and
\Iat has zero average
(e.g. around $\alpha \bar{V}_{\mathrm{g}} \ll T$ it is
 $\Iat \approx ({\Gamma_r} / {\Gamma}) ~ {d p_{t,\downarrow}^{(\mathrm{i})}} / {dt}$).
In general, the interaction breaks the symmetry of loading and unloading parts of the cycle.
The magnetic field, however, restores this symmetry in the linear response regime by lifting the spin degeneracy.
Therefore, \Ia is suppressed for $V_b\ll T$ even though $U\neq 0$.
This is to be contrasted with the standard \didv map shown in the inset,
where in linear response the conductance shows the Coulomb oscillation peaks.
Only at a finite voltage $\bar{V}_\mathrm{b}= B$, where the spin-excited state becomes available,
the loading-unloading symmetry is broken again and \Ia is restored.

{\em Coupling asymmetry.}
Asymmetric rates induce additional features in \Fig{fig:1}(b) and \Fig{fig:1}(d)
at finite bias $|\bar{V}_\mathrm{b}| = U$ and $U-B$, respectively.
In \Fig{fig:2} we plot \IaL along the dashed line in \Fig{fig:1}(a) as function of the coupling asymmetry,
quantified by $\lambda=\left(\Gamma_\mathrm{L}-\Gamma_\mathrm{R}\right)/\Gamma$.
Strikingly,
 the two high bias features marked (3) and (4) are \emph{qualitatively} sensitive to the coupling asymmetry:
if, e.g., resonance (3) is negative (positive), then $\Gamma_\mathrm{L} > \Gamma_\mathrm{R}$ ($\Gamma_\mathrm{L} < \Gamma_\mathrm{R}$).
Quantitatively, for $\lambda > 0$ the adiabatic dc current resonances marked (2) and (3)  deviate from the ``bare'' resonance positions ($\lambda =0$)
by a shift which depends linearly on the temperature $T$~\cite{Deshmukh02,*Golovach04}.
One can thus sensitively probe the coupling asymmetry.

{\em Adiabatic spectroscopy.}
Our results generalize to quantum dots with more complicated states and spectra:
without interaction, the adiabatic dc current vanishes in leading order in $\Gamma$ and $\Omega$.
Therefore, measurement of the time-averaged stability diagrams enables an adiabatic spectroscopy of non-linear transport.
Importantly, the occurrence of adiabatic dc current at sharply defined resonant points,
 indicates that one is measuring in the adiabatic limit.
This relates to the required {effective} two parameter modulation discussed with \Eq{eq:Ia_finiteB}.
Satisfying two SET resonance conditions simultaneously is however not yet sufficient for a non-zero average adiabatic current,
as illustrated above for the crossing of the two ground-to-ground state resonances in a magnetic field.
In general, the occurrence and sign of adiabatic dc current at a charge degeneracy point can be tied to the change in spin-degeneracy in the ground state:
the sign is positive (negative) if the ground state spin-degeneracy increases (decreases) with the quantum dot charge
and it vanishes if there is no change.
The time-averaged stability diagram thus directly reveals non-degenerate ground states
if \IaL vanishes in the linear response regime.
This may be interesting, e.g., for transport through magnetic molecules with high spin degeneracies
or in carbon nanotubes where both spin- and orbital-degeneracies play a role.
Another important aspect of the proposed spectroscopy
is that the effects of ``spurious'' modulation of the barrier can be clearly identified experimentally.
As shown in~\Ref{Splettstoesser06}, a modulation of the gate voltage and of the barrier (instead of the bias voltage) results in an adiabatic dc current which is symmetric with respect to reversal of the time-averaged gate voltage $\alpha \bar{V}_\mathrm{g} \rightarrow U-\alpha \bar{V}_\mathrm{g}$,
in contrast to the antisymmetric shape found here.
The proposed spectroscopy does furthermore not rely on quantum fluctuation effects and can therefore be observed readily in weakly coupled devices at moderate temperature and low driving frequency.
We have checked that the corrections from next-to-leading order tunnel processes ($\Gamma^2$)
to the effects discussed here are quantitative and small, even for $\Gamma \sim T$.
Importantly, even when including these corrections the adiabatic dc current still vanishes exactly for zero interaction.
By measuring the proposed \emph{time-averaged} stability diagram one thus gently probes junction asymmetries and strong interaction effects.
This may prove valuable for instance in molecular quantum dots where stability is a key issue
and transport is the only local probe available.
Adiabatic transport through interacting nano-systems operated in the non-linear regime
is thus a promising topic where new experiments can be done.
We acknowledge
 S. Das, J. K{\"o}nig, M. Plethyukhov, H. Schoeller, C. Stampfer,
and the financial support from
the Helmholtz Foundation,
the FZ-J\"ulich (IFMIT) and the Ministry of Innovation NRW.

\bibliographystyle{apsrev4-1}
\end{document}